\documentclass[twocolumn]{aastex631}
\usepackage[T1]{fontenc}
\usepackage{apjfonts} 

\usepackage{color}
\usepackage{CJKutf8}
\usepackage{subfigure}
\begin{document}
\begin{CJK*}{UTF8}{bsmi}

\newcommand{\vdag}{(v)^\dagger}
\newcommand\aastex{AAS\TeX}
\newcommand\latex{La\TeX}

\newcommand {\msun}{$M_\odot$}
\newcommand {\xray}{X-ray\ }
\newcommand {\ha}{H$\alpha$\ }

\title{Stellar Population near NGC 2021: Procession of Star Formation in the South Rim of Supergiant Shell LMC\,4}

\author[0000-0003-1295-8235]{Po-Sheng Ou （歐柏昇）}
\affiliation{Department of Physics, National Taiwan University, No.1, Sec. 4, Roosevelt Rd.,  Taipei 10617, Taiwan, R.O.C.} 
\affiliation{Institute of Astronomy and Astrophysics, Academia Sinica, No.1, Sec. 4, Roosevelt Rd., Taipei 10617, Taiwan, R.O.C.}

\author{Rui-Ching Chao （趙瑞青）}
\affiliation{Taipei Astronomical Museum, No. 363, Jihe Rd., Taipei 111013, Taiwan, R.O.C.} 

\author[0000-0003-3667-574X]{You-Hua Chu （朱有花）}
\affiliation{Department of Physics, National Sun Yet-Sen University, No.\ 70, Lienhai Rd., Kaohsiung 80424, Taiwan, R.O.C.}
\affiliation{Institute of Astronomy and Astrophysics, Academia Sinica, No.1, Sec. 4, Roosevelt Rd., Taipei 10617, Taiwan, R.O.C.} 

\author{Chin-Yi Hsu （許晉翊）}
\affiliation{Taipei Astronomical Museum, No. 363, Jihe Rd., Taipei 111013, Taiwan, R.O.C.} 

\author[0000-0003-1449-7284]{Chuan-Jui Li （李傳睿）}
\affiliation{Institute of Astronomy and Astrophysics, Academia Sinica, No.1, Sec. 4, Roosevelt Rd., Taipei 10617, Taiwan, R.O.C.}


\begin{abstract}
 Supergiant shells (SGSs) are the largest interstellar structures where 
 heated and enriched gas flows into the host galaxy's halo.  The SGSs in 
 the Large Magellanic Cloud (LMC) are so close that their stars can be resolved 
 with ground-based telescopes to allow studies of star formation history.
 Aiming to study the star formation history and energy budget of LMC\,4, we
 have conducted a pilot study of the cluster NGC\,2021 and the OB associations 
 in its vicinity near the south rim of LMC\,4.  We use the Magellanic Cloud 
 Photometric Survey data of the LMC to establish a methodology to examine
 the stellar population and assess the massive star formation history. 
 We find a radial procession of massive star formation from the northwest 
 part of the OB association LH79 through NGC\,2021 to the OB association LH78 
 in the south.  Using the stellar content of NGC\,2021 and the assumption 
 of Salpeter's initial mass function, we estimate that $\sim$4 supernovae have
 occurred in NGC\,2021, injecting at least $4\times10^{51}$ ergs of kinetic 
 energy into the interior of LMC\,4. 
\end{abstract}

\subjectheadings{ISM: supergiant shell --- ISM: individual objects (LMC\,4) --- Magellanic Clouds}

\section{Introduction}

Supergiant shells (SGSs) are the largest interstellar structures 
excavated by stellar energy feedback in a galaxy.  
Their sizes, $\ga$1000 pc, are usually greater than the scale height 
of the gaseous disk of a galaxy; thus, SGSs are sites where energies 
and enriched material can escape to the galactic halo.  As SGSs 
play an important role in the global evolution of the interstellar 
medium (ISM), their formation mechanism and in particular their 
star formation history (SFH) are of great interest.  

\begin{figure*}[tbh]
\centering
\includegraphics[width=\textwidth]{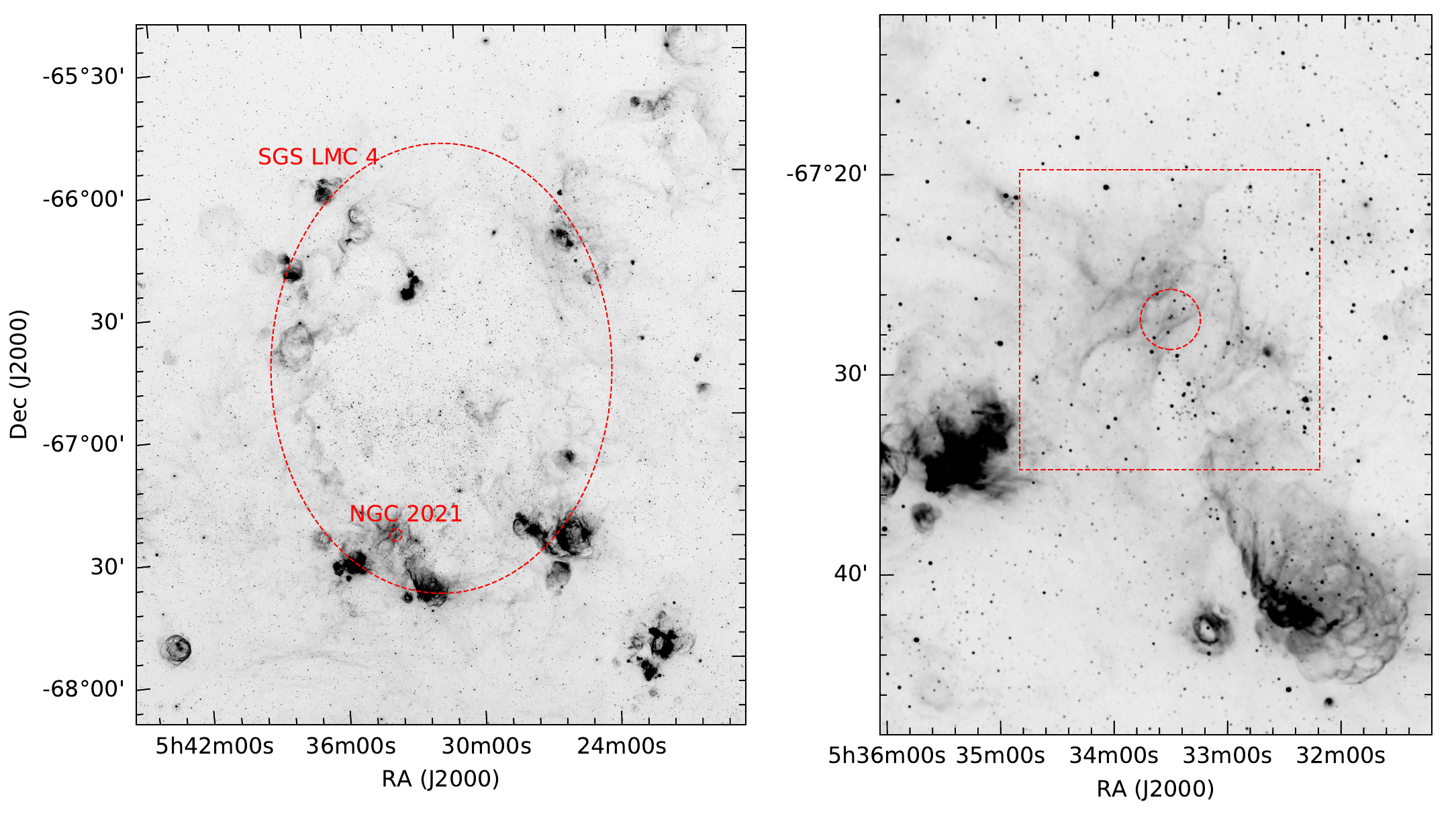}
\caption{Left: MCELS H$\alpha$ image of the SGS LMC\,4, which is 
delineated by the red dashed ellipse. The position of NGC\,2021 is marked on the image.
Right: A close-up of the NGC\,2021 region. The dashed red circle is centered on NGC\,2021, 
and the red square marks the 15$'$$\times$15$'$ field of view of Figure\,\ref{fig:15arcmin}.}
\label{fig:lmc4}
\end{figure*}
\begin{figure*}[tbh]
\centering
\includegraphics[width=15cm]{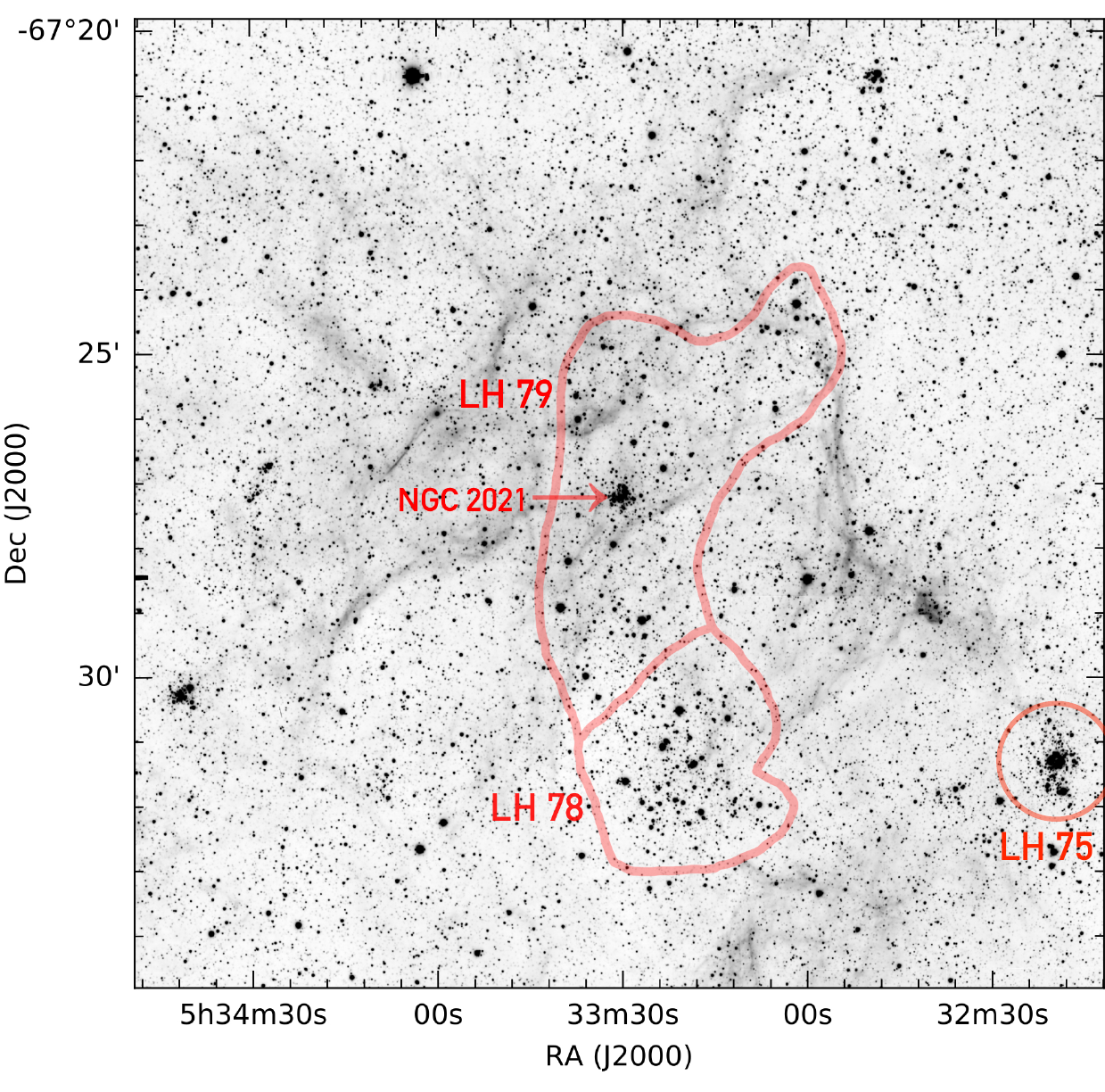}
\caption{CTIO 4\,m MOSAIC [\ion{S}{2}] image of the NGC\,2021 region. 
The [\ion{S}{2}] image is used because both stars and nebulosity can be 
seen in the same image.  The boundaries of the OB associations LH78 and 
LH79 are marked in red according to \citet{Lucke1972}.}
\label{fig:15arcmin}
\end{figure*}

Nine SGSs in the Large Magellanic Cloud (LMC) have been
identified in H$\alpha$ images \citep{Meaburn1980}, although two
of them may not be physical shells \citep{Book2008}.
Among these SGSs, LMC\,4 is the largest, with a dimension of 
1400 pc $\times$ 1000 pc (see Fig.\ \ref{fig:lmc4}). 
The SFH in the SGS LMC\,4 is complex.
Numerous clusters and OB associations are projected within LMC\,4, and
on-going star formation is seen in luminous \ion{H}{2} regions along 
the periphery of LMC\,4.  The stellar populations of selected regions in
LMC\,4 have been studied by various investigators, for example, Shapley
Constellation III \citep{Dolphin1998,Harris2008}, OB associations along 
the rim \citep{Gouliermis2002} or in the interior \citep{Olsen2001} of LMC\,4.  
It has been suggested that stochastic self-propagating
star formation in Shapley Constellation III from the center outwards is
responsible for the formation of LMC\,4 \citep{Feitzinger1981,Dopita1985}; 
however, the youngest stellar populations along the semi-major and near the 
semi-minor axes do not show any age gradient \citep{Braun1997}, challenging the 
self-propagating star formation history \citep{Kamaya1998}.  
To comprehensively investigate the origin and SFH of the SGS LMC\,4, one 
needs to examine the stellar population inside and outside the entire LMC\,4, 
which is made possible by the Magellanic Cloud Photometric
Survey \citep[MCPS;][]{Zaritsky2004} and the Survey of the MAgellanic 
Stellar History \citep[SMASH;][]{Nidever2017}.

The star formation history of the LMC has been studied using the ages of star
clusters determined from the MCPS data \citep{Glatt2010}, Optical Gravitational 
Lensing Experiment II \citep[OGLE II][]{Udalski1997} data \citep{Nayak2016}, 
or Str{\"o}mgren $v$, $b$, and $y$ images obtained with the Southern 
Astrophysical Research (SOAR) Telescope \citep{Narloch2022}.
These studies investigated the formation history of star clusters, but did not
include the field OB stars and thus did not provide a complete picture of
star formation for stellar energy feedback studies.

Before embarking on a thorough investigation of the SFH for the entire 
LMC\,4, including both star clusters and field OB stars,
we conduct a pilot study of the cluster NGC\,2021, which is projected near 
the southeast rim of LMC\,4 (see Fig.\ \ref{fig:lmc4}), in order to 
establish a methodology and assess its limitations. As shown in Figure 2, 
NGC\,2021 is in the central region of the OB association LH79 that 
abuts the OB association LH78 \citep{Lucke1970}.  
We have examined the stellar population
not only in NGC\,2021, LH79, and LH78, but also in the entire 
15$'$$\times$15$'$ (225 pc $\times$225 pc) field in Figure \ref{fig:15arcmin}, 
and analyzed the star formation history.  This paper reports our results.
Section 2 describes the data and methodology we used, Section 3 
reports the stellar population in NGC\,2021, LH78, LH79, and the field,
and Section 4 discusses the star formation history and stellar energy 
feedback in the region we studied.

\section{Available Archival Data and Methodology}

The images used in this paper include H$\alpha$ images from the 
Magellanic Cloud Emission-Line Survey \citep[MCELS;][]{Smith1999}, and
H$\alpha$, [\ion{S}{2}] $\lambda\lambda$6716, 6731, and $R$ band 
images taken with the Blanco 4m Telescope and the MOSAIC camera at 
Cerro Tololo Inter-American Observatory (CTIO). 

The MCPS provides $UBVI$ magnitudes of stars in the Magellanic Clouds (MCs), 
and the limiting magnitudes for completeness are $U=21.5$, 
$B=23.5$, $V=23$, and $I=22$ \citep{Zaritsky2004}.  
As we use massive stars to diagnose star formation history
and the $U$ band is more sensitive to massive stars, we 
use the MCPS data to make $B$ versus ($U-B$) color-magnitude 
diagrams (CMDs).
We initially considered using the SMASH $ugri$ photometric data 
of stars in the LMC; however, the SMASH $u$ band photometry had 
not been accurately calibrated, thus we gave up the SMASH data.

To assess the masses and ages of stars, we plot stellar evolutionary
tracks and isochrones in the CMDs and compare them with the locations 
of stars.  The stellar evolutionary tracks and isochrones are retrieved
from the MESA Isochrones and Stellar Tracks project 
\citep[MIST;][]{Dotter2016,Choi2016}. We use the rotating star models 
with an initial angular velocity to 
critical angular velocity ratio of 0.4 by default. The metallicity is set at 
[Fe/H] = $-$0.37 dex according to the average metallicity of the LMC derived 
from the MCPS data \citep{Choudhury2016}. The extinction of 
each star provided by the MCPS online extinction 
estimator\footnote{\url{https://www.as.arizona.edu/~dennis/lmcext.html}} 
\citep{Zaritsky2004} may have a large error.  Thus, instead of 
dereddening each star, we adopt the average extinction of stars within the
15$'$$\times$15$'$ field centered on NGC\,2021, $A_V=0.48$, and apply it 
to the evolutionary tracks and isochrones.  
The reddened stellar evolutionary tracks of 3, 5, 7, 10, 12, 15, 20, 25, and 
40 $M_{\odot}$ and reddened isochrones for ages of 6, 10, 14, 20, 25, 
and 32 Myr are plotted in the CMDs. 

We use the evolutionary tracks to diagnose massive main-sequence (MS)
stars.  As the $UB$ photometry is not sensitive to the very 
massive stars, we only use the evolutionary tracks to coarsely classify
stars into 3--5, 5--7, 7--10, 10--12, 12--15, 15--20, 20--25, and 
$>$25 $M_\odot$ bins and count the MS stars in these bins.  
Assuming a Salpeter initial mass function \citep{Salpeter1955}, we use 
the star counts to estimate the expected number of massive stars that have exploded 
as supernovae and to assess the stellar energy feedback.  We also use 
the isochrones to assess a lower limit on the stellar populations' ages.

The regions we have analyzed include: \\
(1) NGC\,2021 inside a circular region of 0\farcm5 radius centered at 
 RA = 05$^{\rm h}$33$^{\rm m}$30.12$^{\rm s}$, 
 Dec = $-$67$^{\circ}$27$'$12.9$''$; \\ 
(2) an annular background region for NGC\,2021 with the same center and inner and outer radii of 0\farcm5 to 1\farcm5;\\
(3) a 1\farcm5-radius region in LH79 to the northwest of NGC\,2021 for comparison;\\
(4) a 1\farcm5-radius region centered on LH78; and\\
(5) the entire 15$'$$\times$15$'$ region of Figure~\ref{fig:15arcmin}.

\begin{figure*}[tbh]
\centering
\includegraphics[width=15cm]{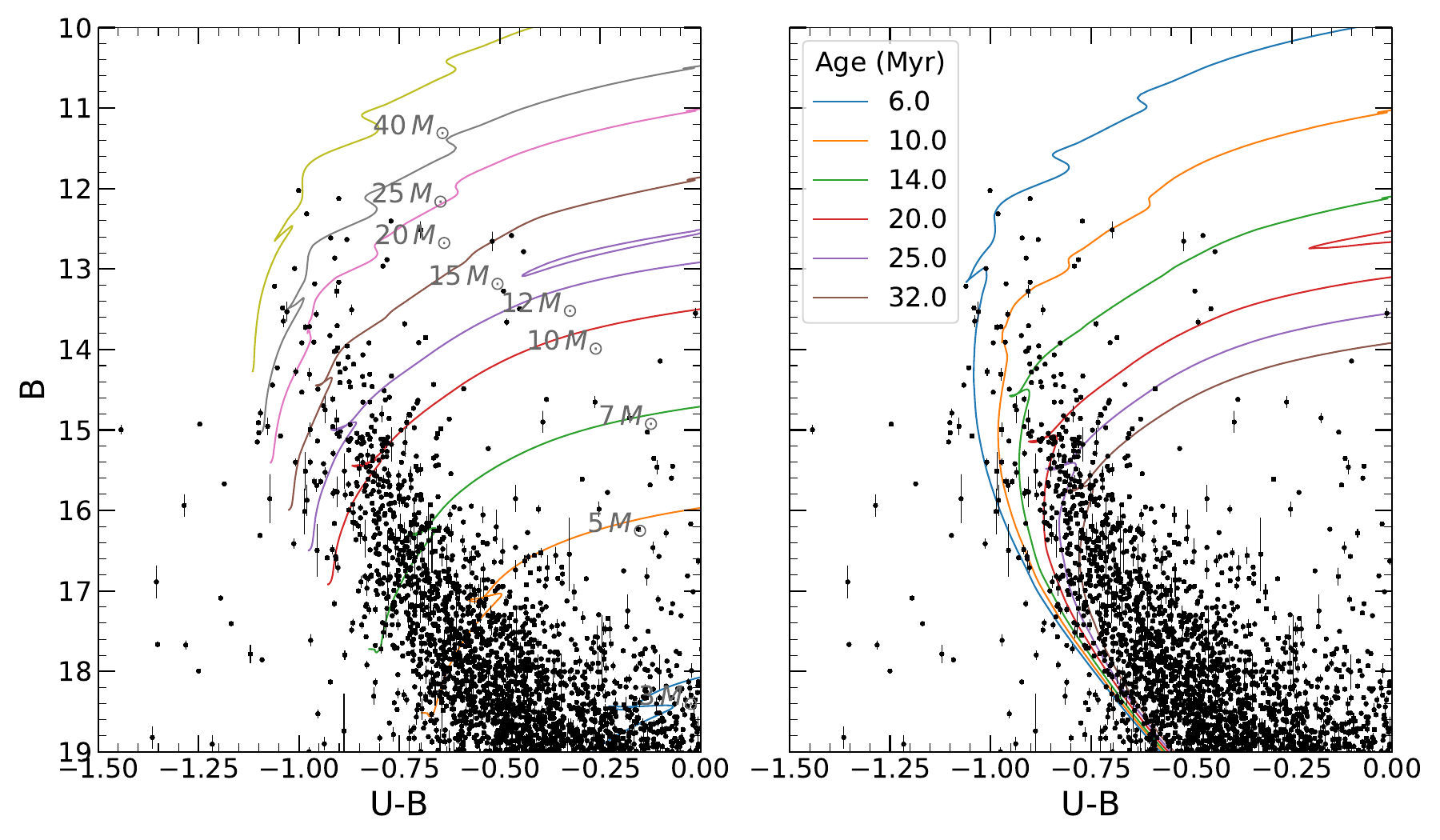}
\caption{The $B$ versus ($U-B$) color-magnitude diagram of stars in the 15$'$$\times$15$'$ 
region centered on NGC\,2021.  The stellar evolutionary tracks and isochrones reddened 
by an extinction of $A_V$ = 0.48 are overplotted in the left and right panels, 
respectively.}
\label{fig:cmd15}
\end{figure*}

\begin{figure*}[tbh]
\centering
\includegraphics[width=15cm]{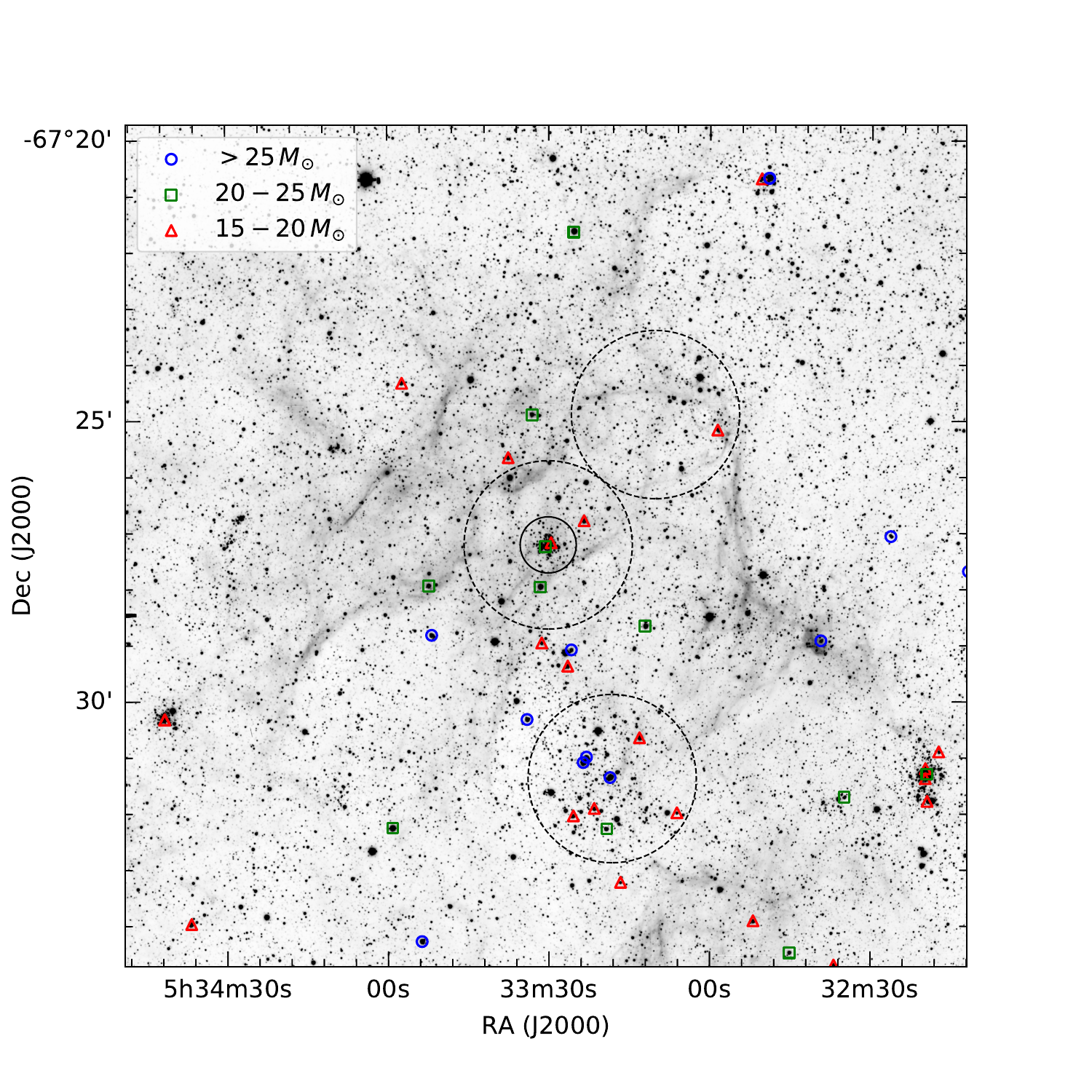}
\caption{CTIO 4\,m MOSAIC [\ion{S}{2}] image of the NGC 2021 region
with massive stars marked. NGC\,2021 is located at the center of the image. The black solid circle shows the NGC\,2021 region with a radius of 0\farcm5. The dashed circles show 
the LH79 NW region, the NGC\,2021 surrounding region, and the LH78 region, all with radii of 1\farcm5.}
\label{fig:LH79-15arcmin}
\end{figure*}

For illustration, in Figure~\ref{fig:cmd15} we present the CMD for Region (5), 
the entire 15$'$$\times$15$'$ field around NGC\,2021.
Comparisons between the locations of stars and the reddened stellar 
evolutionary tracks indicate that no MS stars more massive than 
$\sim$40 $M_\odot$ are present in this field.  Furthermore, comparisons 
between stars and the reddened isochrones indicate that no burst of star 
formation occurred in the past $\sim$6 Myr.
The MS stars identified to have masses $>15\,M_{\odot}$ are
marked on the CTIO 4m MOSAIC [\ion{S}{2}] image in Figure~\ref{fig:LH79-15arcmin}.

\section{Stellar Population in and near NGC\,2021}

As shown in Figure~\ref{fig:15arcmin}, NGC\,2021 is part of
the OB association LH79, which is in juxtaposition to the OB association 
LH78.  We will examine the stellar population in NGC\,2021 and compare
it with that of the northwest portion of LH79.  We then examine LH78 and
compare it with LH79, and further compare them with the 15$'$$\times$15$'$
field.

\subsection{NGC\,2021}

To analyze the stellar population in NGC\,2021, we use a circular region of 0\farcm5 radius 
for the cluster and a concentric annular background region with radii of 
0\farcm5 -- 1\farcm5, as marked in Figure~\ref{fig:LH79-15arcmin}.  
The $B$ versus ($U-B$) CMDs of stellar population in these two regions are presented in 
Figures 5 and 6, respectively.  The reddened stellar evolutionary tracks are overplotted 
in the left panel and the reddened isochrones in the right panel.  
The CMD of NGC\,2021 does not show 
any MS stars with initial masses $\ge$25 $M_\odot$; furthermore, comparisons with the 
isochrones indicate that NGC\,2021 is older than 6 Myr, but not much older than 10 Myr.
The CMD of the background region shows a similar behavior, and thus a similar age.
These results are consistent with the determination of 0-10 Myr age group for 
NGC\,2021 based on integrated $UBV$ photometry \citep{Bica1996}.

To determine the mass function of NGC\,2021, we use the stellar evolutionary
tracks to guide the star counts in stellar mass bins of 3--5, 5--7, 7--10, 10--12, 
and $>$12 $M_\odot$.  The star counts in the annular background region are 
scaled by a factor of 0.5$^2$/(1.5$^2$-0.5$^2$) to account for the surface area 
difference and subtracted from the star counts in the cluster region to produce the 
background-subtracted cluster mass function in Table~\ref{table:num}.

Stars with initial mass estimates are marked in Figure~\ref{fig:LH79-3arcmin}.

\begin{figure*}[tbh]
\centering
\includegraphics[width=15cm]{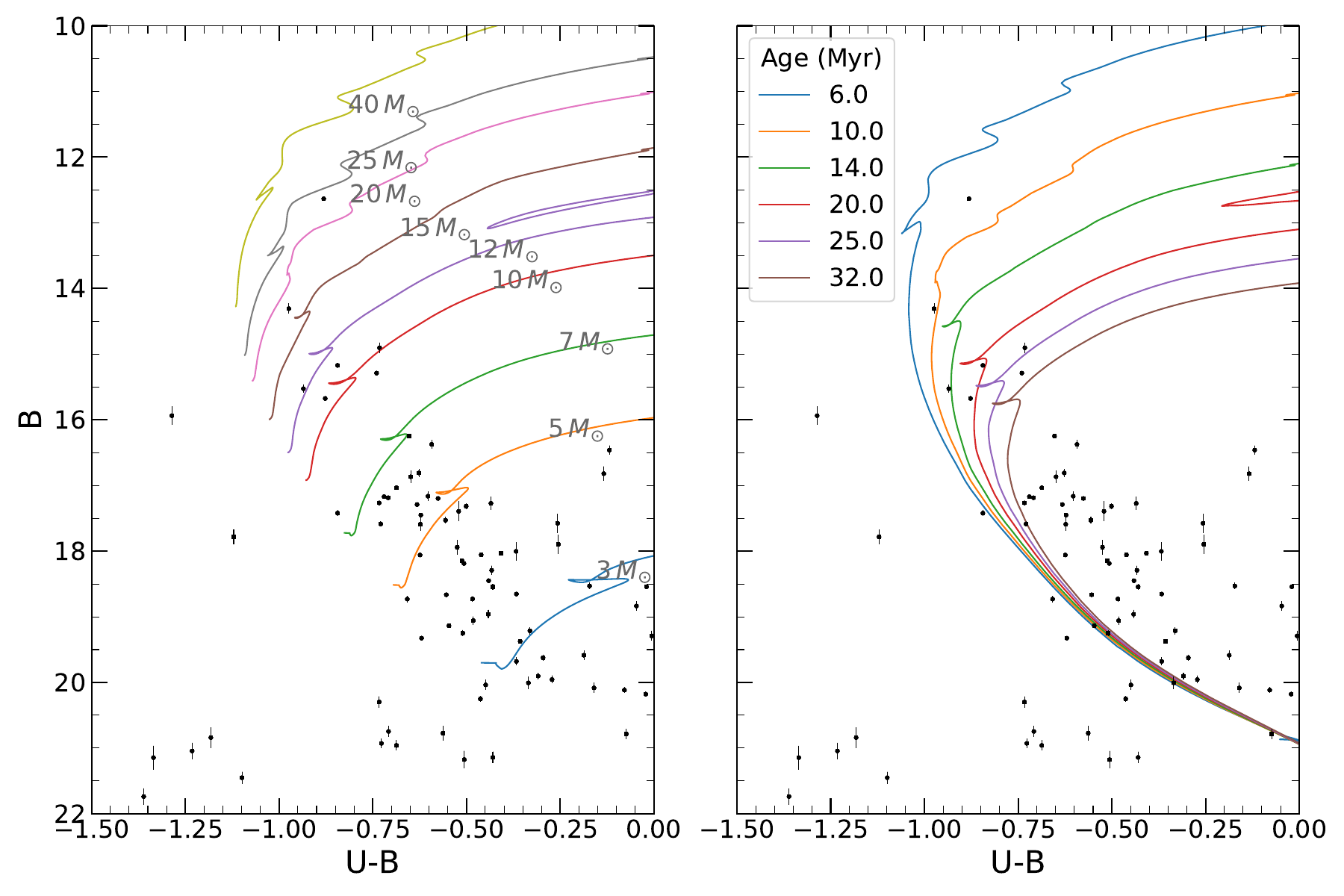}
\caption{Same as Figure\,\ref{fig:cmd15}, but for the region of 0\farcm5 radius 
centered on NGC\,2021.}
\label{fig:cmd1}
\end{figure*}

\begin{figure*}[tbh]
\centering
\includegraphics[width=15cm]{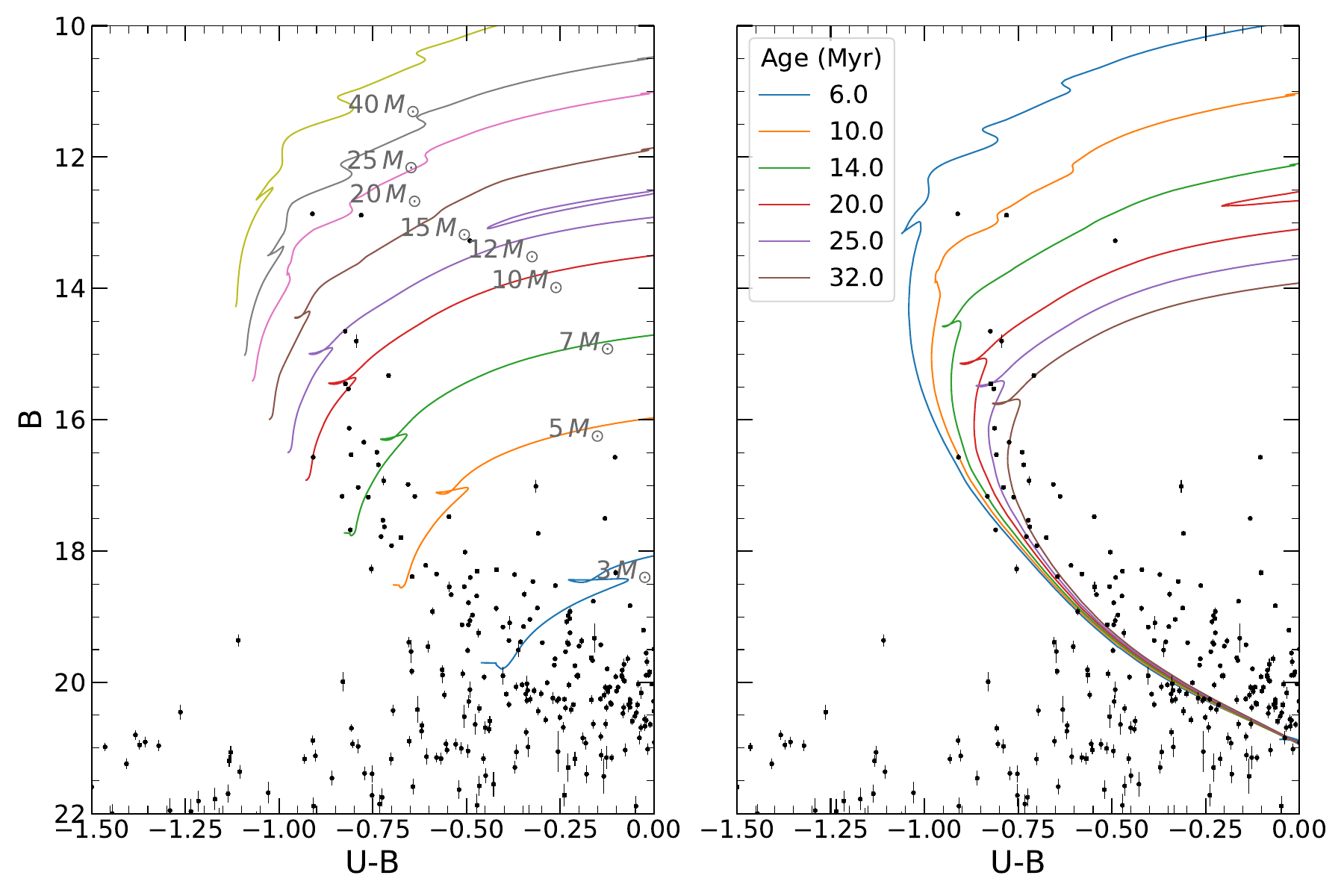}
\caption{Same as Figure\,\ref{fig:cmd15}, but for the annulus of radii 
0\farcm5 -- 1\farcm5 centered on NGC\,2021.}
\label{fig:cmd3}
\end{figure*}

\subsection{LH79-NW}
The stellar population in the northwest part of LH79 is sampled within 
a 1\farcm5-radius region, as marked in Figure~\ref{fig:LH79-15arcmin}.
The CMDs of LH79-NW overlaid with reddened stellar evolutionary tracks and 
isochrones are presented in Figure~\ref{fig:cmdLH79NW}.
No stars with masses $>$20 $M_\odot$ are seen, and comparisons with 
the isochrones indicate the massive stars are older than $\sim$15 Myr, 
with an uncertainty of a few Myr.

\subsection{LH78}

The stars in LH78 are also sampled within a 1\farcm5-radius region, 
as marked in Figure~\ref{fig:LH79-15arcmin}.
The CMDs of LH78 overlaid with reddened stellar evolutionary tracks and 
isochrones are presented in Figure~\ref{fig:cmdLH78}.
LH78 has many bright stars, and its CMD shows a MS populated up to 
25 $M_\odot$ and even evolved higher-mass stars may be present.
Comparisons with isochrones indicate that the massive stars are 
younger than $\sim$6 Myr.

\begin{figure*}[tbh]
\centering
\includegraphics[width=15cm]{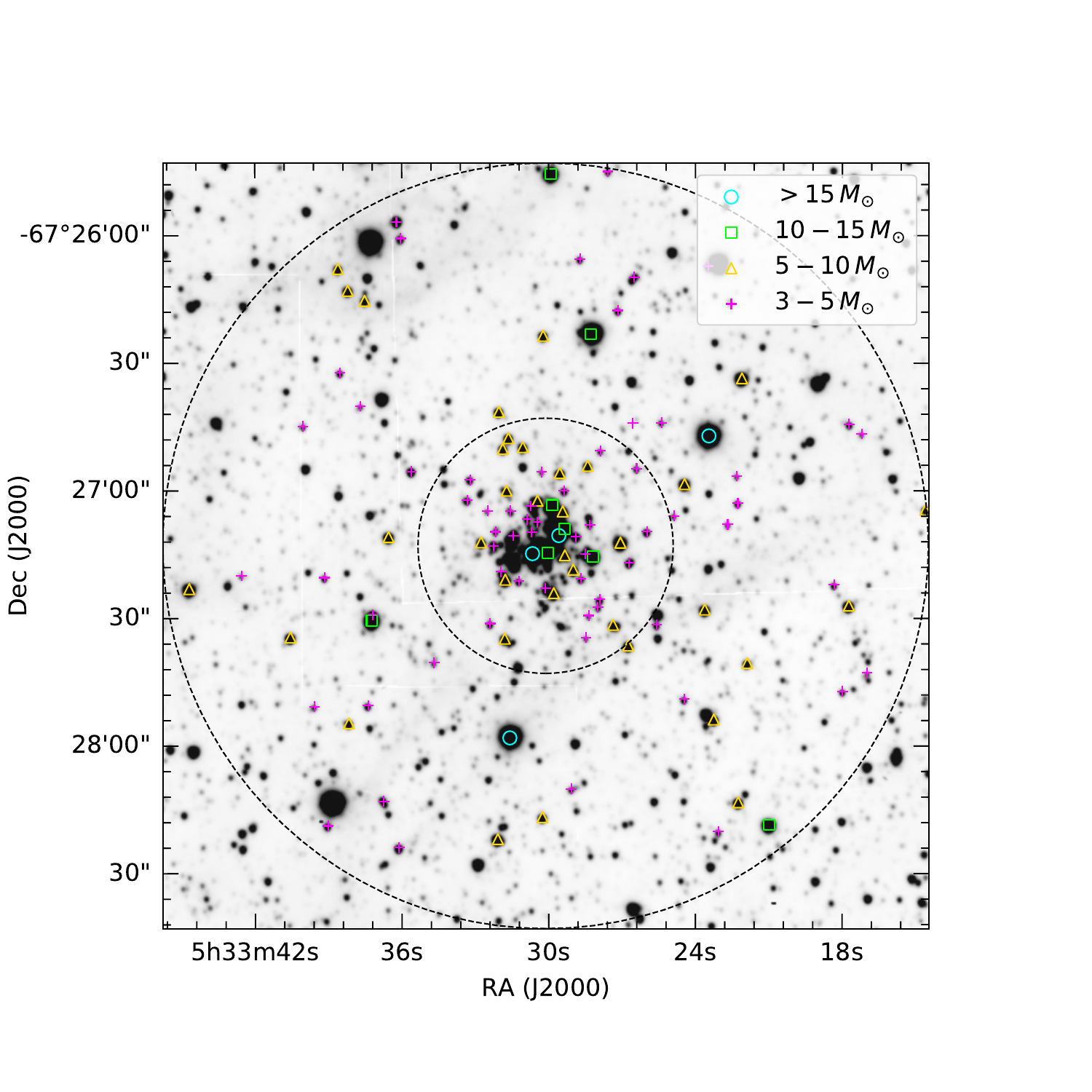}
\caption{In this 4-m MOSAIC R-band image, we present the stellar population 
in NGC\,2021 and its surroundings. The stars with masses of 3--5 $M_{\odot}$ 
(magenta), 5--10 $M_{\odot}$ (yellow), 10--15 $M_{\odot}$ (lime green), 
and $>$15 $M_{\odot}$ (cyan) are marked on the image. The two concentric black 
dashed circles centered on NGC\,2021 are at radii of 0\farcm5 and 1\farcm5, 
respectively.}
\label{fig:LH79-3arcmin}
\end{figure*}

\section{Discussions}

\subsection{Procession of Star Formation near NGC 2021}

The final goal of this work is to probe the formation mechanism of
the SGS LMC 4, and the specific process tested here is whether star 
formation has proceeded radially outward.
The star formation history analyzed in Section 3 show that
the massive stars in LH79-NW are older than $\sim$15 Myr, NGC\,2021
$\sim$6--10 Myr old, and LH78 younger than $\sim$6 Myr.
The sequence of LH79-NW $\rightarrow$ NGC\,2021 $\rightarrow$ LH78 
is not only an age sequence from old to young, but also a radially
outward positional sequence.  Over this length span of about 120 pc, 
recent star formation does seem to have proceeded radially outward.

To examine the distribution of massive stars in a larger field around 
NGC\,2021, we have selected the most massive stars ($>15$ \msun) from 
the CMD of the 15$'$$\times$15$'$ field (Fig.~\ref{fig:cmd15}) and marked 
them in the image in Figure~\ref{fig:LH79-15arcmin}.  It is apparent that stars 
more massive than 25 $M_\odot$ are absent in LH79-NW and NGC\,2021, but
are present in the south end of LH79 and continue into LH78.  The locations
of these massive stars are consistent with the procession of star formation
southward, roughly in the radial direction near the south rim of LMC-4.

\subsection{Stellar Energy Injected by NGC 2021 into the SGS LMC 4}

The mechanical energy injected by NGC\,2021 into the interior of SGS LMC 4
can be assessed from its stellar content.  The background-subtracted 
star counts in different mass ranges listed in Table~\ref{table:num}
represent the present-day mass function (PDMF). 
We assume the \citet{Salpeter1955} initial mass function 
(IMF) $\xi (M) = \xi_0 M^{-2.35}$, where $M$ is the initial 
stellar mass and $\xi_0$ is a scaling factor. 
This scaling factor can be determined from the unevolved low-mass stars 
in the PDMF.  
The difference between the number of stars with initial masses 
$\ge$10 $M_\odot$ observed in the PDMF and that expected from the 
IMF is the number of massive stars that have exploded as supernovae.


First we test whether the numbers of low-mass stars follow the Salpeter IMF.
Denoting the star counts in the initial mass range $M_a$ to $M_b$ as $N_{a-b}$,
the Salpeter IMF requires the ratio of star counts in the $M_1$--$M_2$ and 
$M_3$--$M_4$ ranges to be
$N_{1-2}/N_{3-4} = (M_1^{-1.35}-M_2^{-1.35})/(M_3^{-1.35}-M_4^{-1.35})$.
The observed number ratio of $5-10$ \msun\ and $3-5$ \msun\ stars from 
the observed PDMF is 0.543, which is only $\sim12\%$ off the ratio 
0.612 expected from the Salpeter IMF. We consider these are in reasonable
agreement.
Then we use the number of $3-10$ \msun\ stars and the Salpeter IMF 
to scale the numbers of massive stars (> $10$ \msun). As a result, 
the expected number of stars with initial masses $>10$ \msun\  is 9.3, which is 4.05 larger 
than the number 5.25 derived from the observed data. This result implies 
that 4 massive stars in NGC\,2021 have exploded as supernovae.
Assuming a canonical explosion energy of 10$^{51}$ ergs, NGC\,2021 has 
injected $\sim4\times10^{51}$ ergs of mechanical energy into the SGS LMC 4.
The energy may double, if the mechanical energy of the fast stellar winds
are included.

\begin{deluxetable}{cccc}\label{table:num}
    \tablecaption{Numbers of stars in different mass range}
    \tablehead{Initial mass ($M_{\odot}$) & $N_{\rm{s}}$ & $N_{\rm{bg}}$ & $N$}
\startdata
3-5 & 29 & 35 & 24.625 \\
5-7 & 14 & 10 & 12.75\\
7-10 & 2 & 11 & 0.625\\
10-12 & 3 & 3 & 2.625\\
$>$ 12 & 3 & 3 & 2.625
\enddata
\tablecomments{$N_{\rm s}$ represents the numbers of stars in the cluster NGC\,2021 with a radius of 0\farcm5. $N_{\rm bg}$ represents the numbers of stars in the annular background region between radii of 0\farcm5 and 1\farcm5. $N$ represents the background-subtracted star counts.}
\end{deluxetable}

\begin{figure*}[tbh]
\centering
\includegraphics[width=15cm]{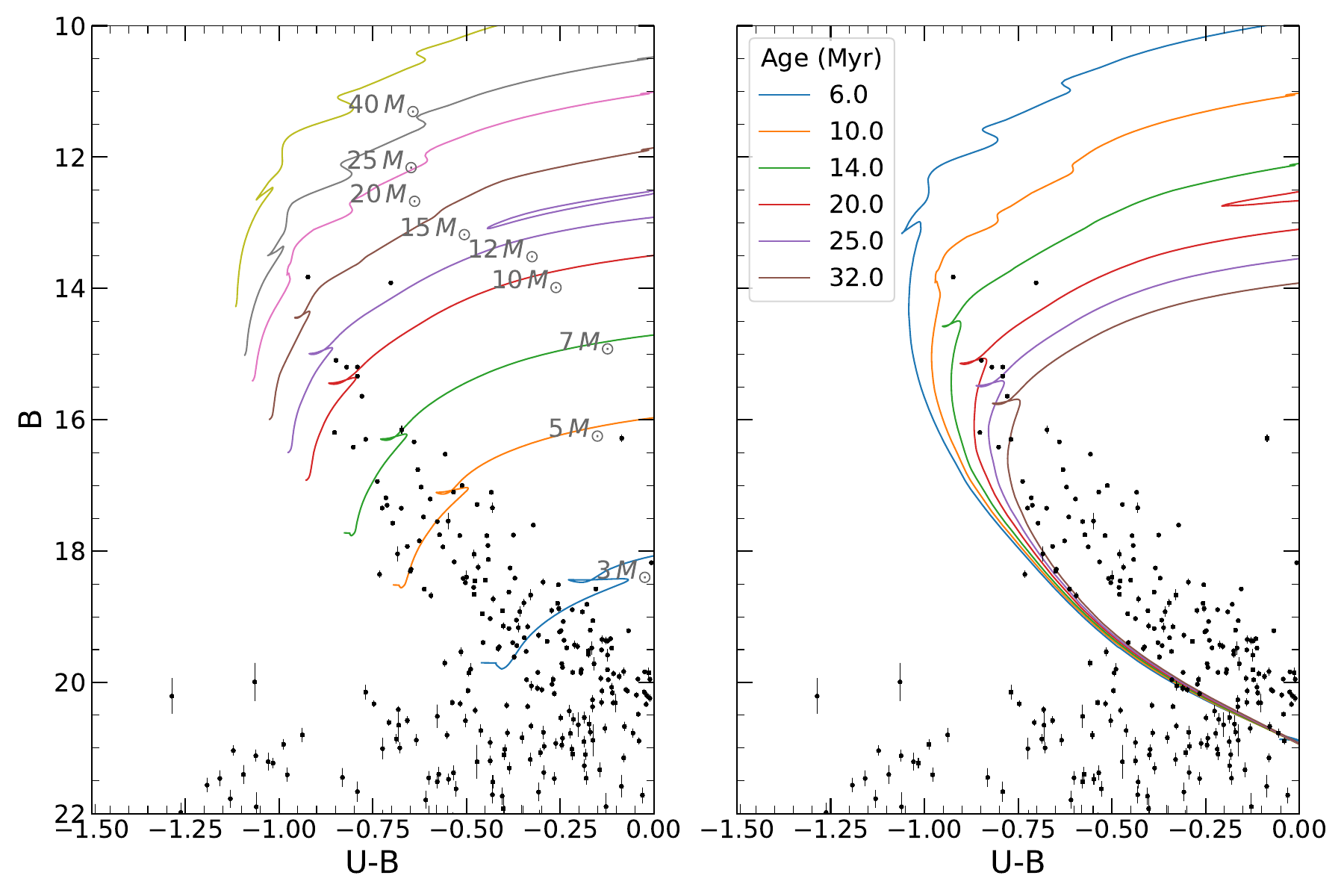}
\caption{Same as Figure\,\ref{fig:cmd15}, but for the LH79-NW region of 1\farcm5 radius.}
\label{fig:cmdLH79NW}
\end{figure*}
\begin{figure*}[tbh]
\centering
\includegraphics[width=15cm]{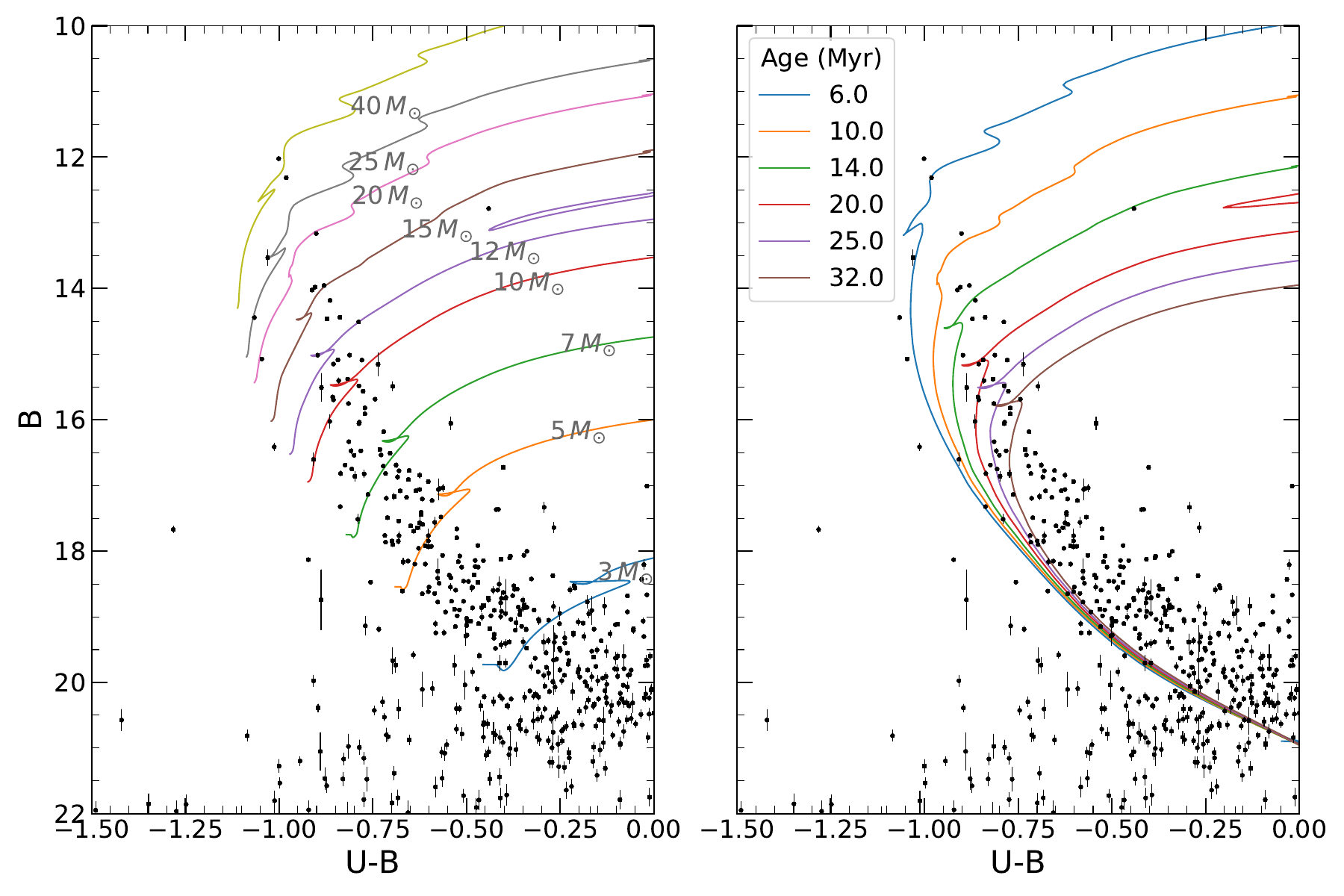}
\caption{Same as Figure\,\ref{fig:cmd15}, but for the LH78 region of 1\farcm5 radius.}
\label{fig:cmdLH78}
\end{figure*}

\section{Summary and Future Plan}

Interstellar gas structures with diameters approaching 1000 pc are
called SGSs.  As their sizes are larger than the host galaxy's
gas disk scale height, they are sites where the heated and enriched gas escapes
into the galactic halo. Recent JWST images reveal that SGSs commonly
exist in disk galaxies \citep[e.g.,][]{Barnes2023}.

The LMC is the only galaxy where the stars in SGSs can be resolved 
and analyzed with ground-based observations.  The largest SGS in the
LMC is LMC 4, and we intend to map its star formation history and
quantitatively assess its energy budget.  As a pilot study, we use
the MCPS $UB$ photometric data of stars in the cluster NGC\,2021 and 
its vicinity near the south rim of LMC 4 to establish a methodology 
to examine the star formation history and stellar energy feedback.

NGC\,2021 is in the OB association LH79. We have made CMDs and compared
the locations of stars with stellar evolutionary tracks and isochrones.
We find an age gradient in the massive stars consistent with a radial
procession of star formation from the north end of LH79 through NGC\,2021
to the OB association LH78.

Our future plan is to recruit high school teachers and students to 
follow the methodology we have established and analyze the massive
star formation history over the entire SGS LMC 4.

\begin{acknowledgments}
This project is supported by the National Science and Technology Council
grant NSTC 111-2112-M-001-063 and 112-2112-M-001-065.
\end{acknowledgments}

\end{CJK*}

\begin{thebibliography}{}

\bibitem[Barnes et al.(2023)]{Barnes2023} 
Barnes, A.~T., Watkins, E.~J., Meidt, S.~E., et al.\ 2023, \apjl, 944, L22. doi:10.3847/2041-8213/aca7b9

\bibitem[Bica et al.(1996)]{Bica1996} 
Bica, E., Claria, J.~J., Dottori, H., et al.\ 1996, \apjs, 102, 57. doi:10.1086/192251

\bibitem[Book et al.(2008)]{Book2008} 
Book, L.~G., Chu, Y.-H., \& Gruendl, R.~A.\ 2008, \apjs, 175, 165. doi:10.1086/523897

\bibitem[Braun et al.(1997)]{Braun1997} 
Braun, J.~M., Bomans, D.~J., Will, J.-M., et al.\ 1997, \aap, 328, 167

\bibitem[Choi et al.(2016)]{Choi2016} Choi, J., Dotter, A., Conroy, C., et al.\ 2016, \apj, 823, 102. doi:10.3847/0004-637X/823/2/102

\bibitem[Choudhury et al.(2016)]{Choudhury2016} Choudhury, S., Subramaniam, A., \& Cole, A.~A.\ 2016, \mnras, 455, 1855. doi:10.1093/mnras/stv2414

\bibitem[Dolphin \& Hunter(1998)]{Dolphin1998} 
Dolphin, A.~E. \& Hunter, D.~A.\ 1998, \aj, 116, 1275. doi:10.1086/300493

\bibitem[Dopita et al.(1985)]{Dopita1985} 
Dopita, M.~A., Mathewson, D.~S., \& Ford, V.~L.\ 1985, \apj, 297, 599. doi:10.1086/163556

\bibitem[Dotter(2016)]{Dotter2016} Dotter, A.\ 2016, \apjs, 222, 8. doi:10.3847/0067-0049/222/1/8

\bibitem[Feitzinger et al.(1981)]{Feitzinger1981} 
Feitzinger, J.~V., Glassgold, A.~E., Gerola, H., et al.\ 1981, \aap, 98, 371

\bibitem[Glatt et al.(2010)]{Glatt2010} Glatt, K., Grebel, E.~K., \& Koch, A.\ 2010, \aap, 517, A50. doi:10.1051/0004-6361/201014187

\bibitem[Gouliermis et al.(2002)]{Gouliermis2002} 
Gouliermis, D., Keller, S.~C., de Boer, K.~S., et al.\ 2002, \aap, 381, 862. doi:10.1051/0004-6361:20011469

\bibitem[Harris \& Zaritsky(2008)]{Harris2008} 
Harris, J. \& Zaritsky, D.\ 2008, PASA, 25, 116. doi:10.1071/AS07037

\bibitem[Kamaya(1998)]{Kamaya1998} 
Kamaya, H.\ 1998, \aj, 116, 1719. doi:10.1086/300543

\bibitem[Lucke(1972)]{Lucke1972} Lucke, P.~B.\ 1972, Ph.D. Thesis

\bibitem[Lucke \& Hodge(1970)]{Lucke1970} 
Lucke, P.~B. \& Hodge, P.~W.\ 1970, \aj, 75, 171. doi:10.1086/110959

\bibitem[Meaburn(1980)]{Meaburn1980} 
Meaburn, J.\ 1980, \mnras, 192, 365. doi:10.1093/mnras/192.3.365

\bibitem[Narloch et al.(2022)]{Narloch2022} Narloch, W., Pietrzy{\'n}ski, G., Gieren, W., et al.\ 2022, \aap, 666, A80. doi:10.1051/0004-6361/202243378

\bibitem[Nayak et al.(2016)]{Nayak2016} Nayak, P.~K., Subramaniam, A., Choudhury, S., et al.\ 2016, \mnras, 463, 1446. doi:10.1093/mnras/stw2043

\bibitem[Nidever et al.(2017)]{Nidever2017} 
Nidever, D.~L., Olsen, K., Walker, A.~R., et al.\ 2017, \aj, 154, 199. doi:10.3847/1538-3881/aa8d1c

\bibitem[Olsen et al.(2001)]{Olsen2001} 
Olsen, K.~A.~G., Kim, S., \& Buss, J.~F.\ 2001, \aj, 121, 3075. doi:10.1086/321092

\bibitem[Salpeter(1955)]{Salpeter1955} Salpeter, E.~E.\ 1955, \apj, 121, 161. doi:10.1086/145971

\bibitem[Smith \& MCELS Team(1999)]{Smith1999} 
Smith, R.~C. \& MCELS Team\ 1999, New Views of the Magellanic Clouds, 190, 28


\bibitem[Udalski et al.(1997)]{Udalski1997} 
Udalski, A., Kubiak, M., \& Szymanski, M.\ 1997, \actaa, 47, 319. doi:10.48550/arXiv.astro-ph/9710091


\bibitem[Zaritsky et al.(2004)]{Zaritsky2004} 
Zaritsky, D., Harris, J., Thompson, I.~B., et al.\ 2004, \aj, 128, 1606. doi:10.1086/423910

\end{thebibliography}
\end{document}